\documentclass{aastex701}

\definecolor{grey}{rgb}{0.9,0.9,0.9}
\usepackage{textgreek}

\begin{document}

\title{Longitudinal distribution of Io's dayside atmospheric density}

\author[orcid=0000-0002-7665-6562]{Rohini S. Giles}
\affiliation{Space Science Division, Southwest Research Institute, San Antonio, Texas, USA}
\email[show]{rohini.giles@swri.org}  

\author[orcid=0000-0003-4452-8109]{John R. Spencer} 
\affiliation{Planetary Science Division, Southwest Research Institute, Boulder, Colorado, USA}
\email{john.spencer@swri.org}

\author[orcid=0000-0001-6613-5731]{Thomas K. Greathouse}
\affiliation{Space Science Division, Southwest Research Institute, San Antonio, Texas, USA}
\email{thomas.greathouse@swri.org}

\author[orcid=0000-0002-1939-6813]{Constantine C. C. Tsang}
\affiliation{Planetary Science Division, Southwest Research Institute, Boulder, Colorado, USA}
\affiliation{Khan Enterprises LLC, Boulder, Colorado, USA}
\email{contsang1981@gmail.com}

\author{Emmanuel Lellouch}
\affiliation{LEISA, Observatoire de Paris/PSL, Meudon, France} 
\email{emmanuel.lellouch@obspm.fr}

\author{Miguel A. López-Valverde}
\affiliation{Instituto de Astrofisica de Andalucia, Granada, Spain}
\email{valverde@iaa.es}

\begin{abstract}

Since 2001, we have used the TEXES high resolution mid-infrared spectrograph at NASA's Infrared Telescope Facility to study the density of Io's SO\textsubscript{2}-dominated atmosphere. The spectra cover several SO\textsubscript{2} absorption lines near 530 cm\textsuperscript{-1} and are used to retrieve the equatorial SO\textsubscript{2} column density. In addition to covering a long time period, our 191 independent observations also cover the full range of Io central longitudes, allowing us to study the longitudinal variability of Io's dayside atmosphere in more depth than has previously been possible. By improving our modeling of longitudinal variability and using a seasonal model to apply a temporal normalization to the retrieved SO\textsubscript{2} densities, we produce two complete longitudinal distributions of Io's equatorial SO\textsubscript{2} column abundances, one representing perihelion and one representing aphelion. Both show the same hemispheric asymmetry that has been observed in previous studies, with the highest densities near 200$^{\circ}$W and the lowest densities near 305$^{\circ}$W. We find that the contrast between the maximum and minimum densities is a factor of $\sim$27 at aphelion and a factor of $\sim$16 at perihelion. Both longitudinal distributions are asymmetric, with a sharper drop-off on Io's trailing hemisphere than on the leading. We compare these longitudinal distributions to the distributions expected if the spatial variability were primarily driven by volcanic activity, SO\textsubscript{2}-ice inhomogeneity or the effects of eclipses. We conclude that the effect of eclipses is important, but insufficient to explain the shape of the observed longitudinal variability, and frost distribution is likely to also be important.

\end{abstract}


\section{Introduction}
\label{sec:introduction}

The density of Io's SO\textsubscript{2}-dominated atmosphere is highly variable, both spatially and temporally. The atmosphere is mostly limited to the equatorial regions of the moon, with densities dropping off rapidly at latitudes greater than 40$^{\circ}$~\citep{jessup04,feaga09,depater20}. The atmosphere partially collapses when the moon enters a Jupiter eclipse, reforming when it re-enters sunlight~\citep{retherford07,tsang16,depater20}. Decades of monitoring have also shown that there is a strong seasonal component, with higher densities when Io is closer to the Sun and lower densities when it is further away~\citep{tsang12,giles24}. In this paper, we focus on the longitudinal variability of Io's dayside atmosphere.

\cite{jessup04} were the first to discuss the possibility of significant longitudinal variability in the density of Io's dayside SO\textsubscript{2} atmosphere. They used the Hubble Space Telescope Imaging Spectrograph (HST/STIS) to obtain spatially resolved ultraviolet (UV) spectra of Io's anti-Jovian hemisphere, and used these to retrieve the SO\textsubscript{2} column density. They found that the values obtained on the anti-Jovian hemisphere were eight times higher than the densities obtained five years earlier from UV observations of the trailing hemisphere~\citep{mcgrath00}, despite the earlier observations being centered on a volcanic plume region. Rather than attributing this to temporal variability, \cite{jessup04} suggested that this apparent discrepancy could represent a genuine longitudinal variability, perhaps driven by differences in the hemispheric SO\textsubscript{2} frost abundances.

The longitudinal variability in the dayside atmosphere was then confirmed by mid-infrared observations obtained by~\cite{spencer05} using TEXES, a high-resolution mid-infrared spectrograph at NASA's Infrared Telescope Facility (IRTF). Over three observing runs in November 2001, December 2001 and January 2004,~\cite{spencer05} systematically obtained a series of disk-averaged observations of Io, each with a different Io-Jupiter geometry and hence a different Io central longitude. The 530 cm\textsuperscript{-1} SO\textsubscript{2} absorption features captured in these high resolution spectra showed significant differences in depth depending on the hemisphere that was observed, even for observations obtained during the same observing run, indicative of large longitudinal differences in the column density of SO\textsubscript{2}. The spectra were modeled by combining the results of a thermal emission model with the calculated transmission of the atmosphere for a given SO\textsubscript{2} column density, and then integrating across the disk using an assumed latitudinal distribution. The best-fit equatorial SO\textsubscript{2} column density was retrieved for each independent observation using a \textchi\textsuperscript{2}-minimization technique, and by combining observations obtained at different central longitudes, they were able to obtain a longitudinal distribution. They found that the maximum column density of 1.5$\times$10\textsuperscript{17} cm\textsuperscript{-2} occurred on the anti-Jovian hemisphere, while the minimum density of 1.5$\times$10\textsuperscript{16} cm\textsuperscript{-2} occured on the sub-Jovian hemisphere at $\sim$300$^{\circ}$W.

Longitudinal variability in Io's atmospheric density has also been studied using ultraviolet observations from the Hubble Space Telescope (HST). \cite{feaga09} used spatially-resolved Ly-\textalpha~images of Io to infer a factor of 5 difference in SO\textsubscript{2} column density between the densest equatorial longitudes near the anti-Jovian point and the least dense equatorial longitudes near the sub-Jovian point. ~\cite{tsang13} obtained a similar result using a series of six disk-averaged mid-UV spectra, evenly spaced across Io central longitudes. Further confirmation of the hemispheric asymmetry has also been provided by 4-\textmu m spectroscopy~\citep{lellouch15}.

There have been several theories put forward about what drives the longitudinal inhomogeneity of Io's atmosphere. Io's atmosphere is thought to be largely supported by sublimation of the surface SO\textsubscript{2} ice, with volcanism potentially playing a significant but smaller role~\citep{depater23}. \cite{feaga09} have suggested that the longitudinal variability may be caused by the non-uniform distribution of volcanoes across Io's surface, while~\cite{jessup04} and \cite{spencer05} proposed that the primary factor is the spatial variability of SO\textsubscript{2} ice. For a sublimation-supported atmosphere, the eclipses that occur when Jupiter passes between Io and the Sun are also expected to contribute to the hemispheric asymmetry, by lowering the daytime temperatures and SO\textsubscript{2} vapor pressures on the sub-Jovian hemisphere relative to the anti-Jovian hemisphere~\citep{tsang12,walker12,dott25}.

An accurate, complete longitudinal distribution is key to distinguishing between these different possible mechanisms. Ly-\textalpha~imaging provides excellent spatial resolution, with the distinct advantage of being able to measure latitudinal as well and longitudinal variability~\citep{feaga09}. However, ~\cite{giono21} have found that these images are only sensitive to SO\textsubscript{2} column densities of 1$\times$10\textsuperscript{15}--5$\times$10\textsuperscript{16} cm\textsuperscript{-2}, lower than the peak equatorial abundances observed in the mid-infrared and ultraviolet~\citep{spencer05,tsang13}. Here, we make use of ground-based mid-infrared observations to seek to improve our understanding of Io's longitudinal atmospheric variability.

For the past 25 years, we have been regularly obtaining observations of Io using the TEXES spectrograph at the IRTF, using the same 530 cm\textsuperscript{-1} spectral set-up used in~\cite{spencer05}. This long-term dataset has been used to track the seasonal variability of both the anti-Jovian and sub-Jovian hemispheres over the course of two Jovian years~\citep{giles24}, confirming the importance of sublimation in supporting Io's atmosphere. In addition to spanning a large timeframe, this dataset also spans a large range of Io central longitudes, providing the opportunity to derive a more complete longitudinal distribution than has previously been possible. 

In Sections~\ref{sec:observations} and~\ref{sec:data_reduction} we describe the TEXES observations and the data reduction process. In Section~\ref{sec:spectral} we discuss how the data is used to retrieve the SO\textsubscript{2} column abundances, including new improvements we have made to the model to account for the longitudinal variability when calculating the disk-averaged spectra. The updated version of the seasonal variability model is then presented in Section~\ref{sec:seasonal}. In Section~\ref{sec:longitudinal}, we use the seasonal model to apply a temporal normalization to the longitudinal data,  allowing us to combine observations from slightly different heliocentric distances into a single longitudinal distribution plot. We use this method to produce two longitudinal distributions of Io's equatorial SO\textsubscript{2} column abundances, one representing perihelion and one representing aphelion. In Section~\ref{sec:results}, these results are discussed and compared to the expected distributions due to volcanic activity, SO\textsubscript{2} frost sublimation, and eclipses.

\begin{longrotatetable}
\begin{deluxetable}{ccccc}
\tabletypesize{\footnotesize}
\tablewidth{0pt}
\tablecaption{Summary of IRTF observing runs contributing to this study \label{tab:run_summary}}
\tablehead{
\colhead{Start date} & \colhead{End date} & \colhead{Number of observations} & \colhead{Io central longitudes ($^{\circ}$W)} & \colhead{Heliocentric distance (AU)}
}
\startdata
2001-11-16 & 2001-11-18 & 3 & 16, 38, 213 & 5.15 \\
2002-12-06 & 2002-12-19 & 7 & 23, 33, 44, 93, 196, 214, 321 & 5.30 \\
2004-01-06 & 2004-01-08 & 3 & 99, 139, 153 & 5.41 \\
2005-01-16 & 2005-01-19 & 4 & 104, 156, 268, 306 & 5.45 \\
2007-04-14 & 2007-04-18 & 6 & 68, 78, 165, 214, 323, 331 & 5.34 \\
2009-06-02 & 2009-06-07 & 9 & 107, 116, 124, 156, 218, 225, 265, 273, 282 & 5.06 \\
2010-06-02 & 2010-06-03 & 3 & 158, 164, 317 & 4.97 \\
2012-01-08 & 2012-01-17 & 10 & 90, 98, 106, 111, 129, 133, 137, 145, 152, 311 & 4.97 \\
2012-10-07 & 2012-10-11 & 8 & 43, 50, 210, 221, 231, 294, 303, 314 & 5.04 \\
2013-02-04 & 2013-02-12 & 12 & 103, 110, 124, 133, 144, 153, 237, 245, 283, 293, 325, 336 & 5.08 \\
2014-02-20 & 2014-02-20 & 1 & 203 & 5.21 \\
2014-12-07 & 2014-12-12 & 4 & 157, 165, 201, 210 & 5.31 \\
2015-11-15 & 2015-11-15 & 2& 138, 148 & 5.41 \\
2016-04-30 & 2016-04-30 & 2 & 198, 210 & 5.44 \\
2018-02-10 & 2018-02-10 & 1 & 203 & 5.43 \\
2018-07-19 & 2018-07-19 & 1 & 98 & 5.39 \\
2019-04-15 & 2019-04-23 & 8 & 59, 154, 213, 221, 260, 269, 324, 331 & 5.32 \\
2019-08-15 & 2019-08-20 & 4 & 224, 272, 281, 290 & 5.28 \\
2021-07-04 & 2021-07-07 & 12 & 56, 63, 79, 103, 262, 270, 277, 283, 310, 318, 326, 332 & 5.04 \\
2021-09-27 & 2021-09-29 & 3 & 52, 98, 106 & 5.01 \\
2021-11-06 & 2021-11-09 & 8 & 74, 83, 91, 97, 122, 130, 234, 243 & 5.00 \\
2022-06-26 & 2022-07-03 & 13 & 100, 108, 200, 209, 218, 248, 256, 297, 304, 312, 313, 320, 330 & 4.96 \\
2022-09-15 & 2022-09-18 & 12 & 73, 80, 88, 114, 122, 130, 230, 238, 246, 265, 273, 280 & 4.96 \\
2023-07-20 & 2023-07-25 & 14 & 75, 83, 118, 126, 133, 165, 232, 242, 277, 285, 291, 321, 329, 336 & 4.96 \\
2023-10-01 & 2023-10-02 & 4 & 129, 335, 343, 326 & 4.97 \\
2023-12-07 & 2023-12-11 & 10 & 85, 93, 134, 141, 148, 242, 250, 257, 315, 322 & 4.98 \\
2024-02-23 & 2024-02-26 & 5 & 87, 135, 142, 243, 250 & 5.00 \\
2025-02-28 & 2025-03-06 & 11 & 121, 129, 135, 214, 222, 259, 267, 274, 323, 331, 338 & 5.10 \\
2025-09-20 & 2025-09-23 & 8 & 106, 114, 148, 157, 260, 270, 303, 312 & 5.17 \\
2026-03-04 & 2026-03-05 & 3 & 152, 162, 301 & 5.23 \\
\enddata
\end{deluxetable}
\end{longrotatetable}

\section{Analysis}
\label{sec:analysis}

\subsection{TEXES observations}
\label{sec:observations}

Since 2001, the TEXES mid-infrared spectrograph~\citep{lacy02} at NASA's Infrared Telescope Facility has been regularly used to obtain observations of SO\textsubscript{2} absorption in Io's atmosphere. To date, there have been 30 observing runs, summarized in Table~\ref{tab:run_summary}. This long-term dataset has previously been analyzed in ~\cite{spencer05},~\cite{tsang12,tsang13b} and~\cite{giles24}. The observations, data reduction and spectral modeling follow the same methods described in the earlier papers, with the exception of the improvement in the longitudinal modeling described in Section~\ref{sec:model_improvement}, and include an additional six observing runs from 2023--2026.

\begin{figure*}
\centering
\includegraphics[width=16cm]{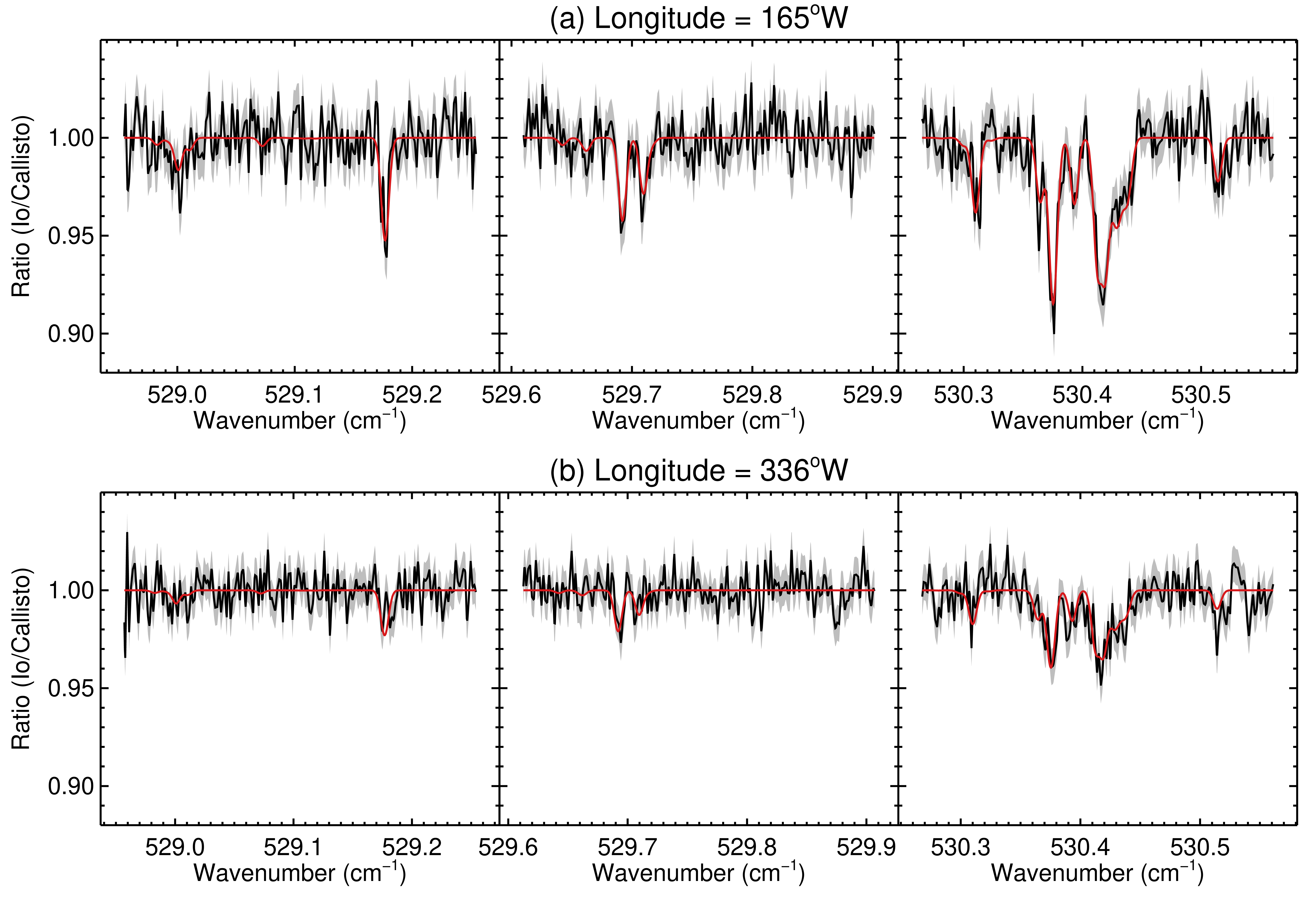}
\caption{Two reduced Io spectra, showing the difference in the SO\textsubscript{2} absorption features between the anti-Jovian hemisphere (central longitude of 165$^{\circ}$W) and the sub-Jovian hemisphere (central longitude of 336$^{\circ}$W). The data were obtained on two consecutive nights (July 25 2023 for (a) and July 24 2023 for (b)). The data are shown in black with the noise in gray. The spectra have each been divided by a Callisto spectrum, flattened and normalized. The corresponding best-fit model spectrum is shown in red.}
\label{fig:spectrum}
\end{figure*}

TEXES was used in its high resolution mode, which provides a resolving power of 75,000. The observations were centered at 529.8 cm\textsuperscript{-1} (corrected for Io's Doppler shift) and have a bandpass of $\sim$1.6 cm\textsuperscript{-1} across three cross-dispersed orders. The wavelength coverage of this set-up can be seen in the spectra shown in Figure~\ref{fig:spectrum}. This spectral set-up captures a strong SO\textsubscript{2} absorption feature at 530.4 cm\textsuperscript{-1}, as well as several weaker lines. Figures~\ref{fig:spectrum}(a) and ~\ref{fig:spectrum}(b) show data obtained on two consecutive nights, from the anti-Jovian and sub-Jovian hemispheres of Io respectively. The data show a clear difference in the depth of the SO\textsubscript{2} absorption features, demonstrating the strong longitudinal variations in the SO\textsubscript{2} column density. 

The observations were obtained by nodding Io 6'' along the TEXES instrument slit; this allows for background subtraction, while maximizing integration time on source. The 2'' instrument slit is wider than the angular diameter of Io, which ranges between 0.8'' and 1.3'', so the data obtained is a disk-average of one hemisphere, centered at a given Io longitude. By obtaining multiple $\sim$40-min observations over the course of several nights, with different Io-Jupiter geometries, we are able to build up longitudinal coverage. The central longitudes of each independent observation are listed in Table~\ref{tab:run_summary}. At these wavelengths, Callisto is bright and spectrally gray. We therefore obtained comparable $\sim$5-min nodded observations of Callisto before and after each Io observation, to act as a divisor for removal of instrumental and telluric effects.

\subsection{Data reduction}
\label{sec:data_reduction}

The Io and Callisto observations were reduced using the standard TEXES data reduction pipeline~\citep{lacy02}. This pipeline performs flat fielding and sky subtraction, makes initial corrections for instrumental distortions and uses telluric absorption lines to calibrate wavelengths. The data from each 40-min Io observing sequence and each 5-min Callisto observing sequence were then co-added by weighting each nod-pair by the square of the signal-to-noise. The Io spectra were then divided by the Callisto spectra with the closest airmass, to remove remaining instrumental effects and telluric features. For each divided spectrum, a curve was fitted to the segments of the spectrum that are free from SO\textsubscript{2}-absorption and the spectrum was divided by this curve, correcting any residual slopes and normalizing the continuum to 1. Finally, the spectra were Doppler-shifted into Io's rest frame. The black lines in Figure~\ref{fig:spectrum} show the final reduced spectra from two observations made in July 2023. 

\subsection{Spectral modeling}
\label{sec:spectral}

\subsubsection{Initial model}

The spectral model used to retrieve the equatorial SO\textsubscript{2} column densities from the absorption spectra was previously described in ~\cite{spencer05},~\cite{tsang12} and~\cite{giles24}. The SO\textsubscript{2} absorption lines at 529--531 cm\textsuperscript{-1} are due to the first excited state of the \textnu\textsubscript{2} mode of vibration. As Io's atmosphere is very low density and is not in local thermodynamic equilibrium (LTE), a non-LTE model was used to calculate the population of this excited state. This model was then combined with a surface thermal emission model from~\cite{veeder94} to generate a synthetic top-of-atmosphere spectrum for a given SO\textsubscript{2} column density. To calculate a synthetic disk-averaged spectrum for a specified equatorial SO\textsubscript{2} column density, we then integrated across the disk, taking into account the change in emission angles and assuming that the SO\textsubscript{2} column density decreases away from the equator according to the ``modified latitude'' model of~\cite{spencer05}, which approximates the relative latitudinal distribution derived in~\cite{jessup04}.

This model assumes a fixed atmospheric temperature of 108 K, based on~\cite{tsang12}. In~\cite{giles24}, we discussed how the lower atmospheric temperatures required to fit the mid-infrared spectra are not consistent with the much higher temperatures retrieved from mm-wave and near-infrared spectra. We note now that \cite{proudkii26} have recently inferred the first vertically-resolved temperature profiles of Io's atmosphere, and their finding that the lower atmosphere is relatively cold (124–-137 K), followed by a thermospheric rise to several hundred Kelvin, may explain this apparent discrepancy.

Model spectra were generated for a range of SO\textsubscript{2} column densities and were smoothed to match the spectral resolution of the TEXES observations. For each of the 191 independent observations listed in Table~\ref{tab:run_summary}, the best-fit SO\textsubscript{2} equatorial column density was retrieved by using a \textchi\textsuperscript{2}-minimization code to compare the data with the model spectra. The red lines in Figure~\ref{fig:spectrum} show the best-fit model spectra for the two observations obtained at different sub-observer longitudes, and correspond to an equatorial SO\textsubscript{2} column density of 1.64$\times$10\textsuperscript{17} cm\textsuperscript{-2} for Figure~\ref{fig:spectrum}(a) and an equatorial SO\textsubscript{2} column density of  0.58$\times$10\textsuperscript{17} cm\textsuperscript{-2} for Figure~\ref{fig:spectrum}(b).

The errors on these retrieved SO\textsubscript{2} abundances were calculated following the same method described in~\cite{giles24}, taking into account both the error in the assumed atmospheric temperature~\citep[12 K,][]{tsang12} and the noise on the observed data. A synthetic `observed' spectrum was produced by adding an appropriate level of randomly-generated observational noise to the best-fit model spectrum, and it was then fitted using the same method as the real spectrum, using an atmospheric temperature that was randomly perturbed about the mean 108-K value. This process was repeated 1000 times, and the standard deviation of these SO\textsubscript{2} column densities was taken as the error on the best-fit SO\textsubscript{2} column density.

While the forward model that generated the synthetic disk-integrated spectra did take into account latitudinal variability in the SO\textsubscript{2} density, using the distribution from~\cite{jessup04}, it is important to note that this initial model used in~\cite{spencer05},~\cite{tsang12} and~\cite{giles24} did not take into account any longitudinal variability. For a given disk-integrated observation, the sub-observer equatorial SO\textsubscript{2} column density was assumed to be constant with longitude, although observations have shown that there is in fact considerable longitudinal variability. As noted in~\cite{giles24}, this simplification would have caused the inferred longitudinal variability derived by combining multiple disk-averaged observations~\citep{spencer05,tsang12,giles24}, to be blurred. In Section~\ref{sec:model_improvement}, we describe the changes we make to this initial model in order to account for longitudinal variability. The model also does not account for any diurnal variability, i.e. the effect of varying amounts of incident solar radiation between the center of the sunlit disk and the edges of the disk. However, unlike the longitudinal variability, previous studies have shown that there is minimal diurnal variability across the disk~\citep{feaga09,jessup15} so we keep this assumption. The model also continues to assumes a constant atmospheric temperature across all longitudes, as altering the atmospheric temperature does not significantly affect the quality of the fits~\citep{tsang12}.

\subsubsection{Improved model}
\label{sec:model_improvement}

In order to account for longitudinal variability in the spectral forward model, we first need to define a subset of our IRTF/TEXES observations to use in the process. We are seeking data that provide broad longitudinal coverage but can be directly compared without taking into account seasonal effects. \cite{giles24} showed that there was significant seasonal variability but minimal inter-annual variability, allowing us to combine data obtained approximately one Jovian year apart. We have obtained more data close to perihelion than at any other season, so in this section, we restrict our subset of data to time periods that are close to perihelion; using the modeled seasonal curve shown in Figure 3 of \cite{giles24}, we use dates where the modeled anti-Jovian SO\textsubscript{2} column density is 2.0$\pm$0.1$\times$10\textsuperscript{17} cm\textsuperscript{-2}. This gives us 63 independent observations, from observing runs carried out in January 2012, June 2022, September 2022, July 2023, September 2023 and December 2023.

\begin{figure*}
\centering
\includegraphics[width=11cm]{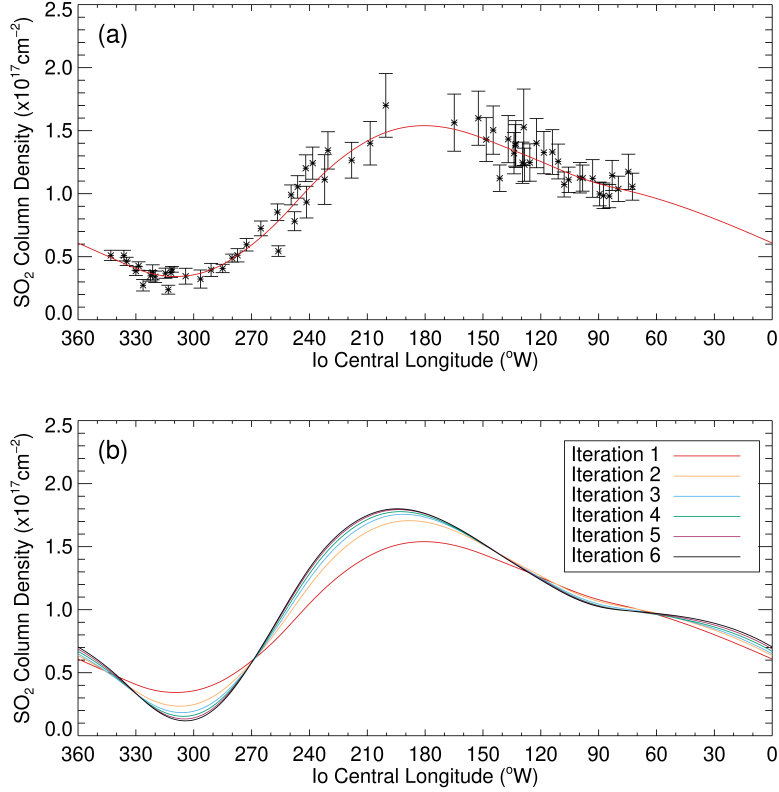}
\caption{(a) Equatorial SO\textsubscript{2} column density as a function of central longitude for the 63 independent observations described in Section~\ref{sec:model_improvement}. These values were retrieved using the former version of the forward model, which did not take into account any longitudinal variability across the disk. The red line shows a fit to the data using a cubic spline, and this fitted curve was the longitudinal distribution used in Iteration 1 of the model improvement. (b) The longitudinal distributions produced at each step of the iterative model improvement. The black line, showing the results of Iteration 6, is the longitudinal distribution used in the final version of the forward model.}
\label{fig:iteration}
\end{figure*}

Using these data, we perform an iterative process to converge on a longitudinal distribution that will be used in the final version of the forward model. The black data points in Figure~\ref{fig:iteration}(a) shows the retrieved SO\textsubscript{2} column densities from these 63 observations, using the initial version of the forward model, before any longitudinal variability in the atmospheric density is taken into account in the model, i.e. a longitudinally-flat distribution is assumed. We take these data points and use a cubic spline fit to produce a continuous longitudinal distribution, which will serve as the assumed longitudinal distribution for the next iteration of the forward model.

In this new version of the forward model, we add the sub-observer longitude as a second variable when generating the set of synthetic disk-integrated spectra, along with the sub-observer SO\textsubscript{2} column density. When calculating the disk-integrated spectrum for a given SO\textsubscript{2} column density and longitude, the equatorial SO\textsubscript{2} abundances are assumed to follow the longitudinal distribution defined by the previous iteration, scaled to match the given SO\textsubscript{2} density at the sub-observer longitude.

Using this new forward model, we then re-retrieve the SO\textsubscript{2} column densities for the 63 data points, and we repeat this process until we converge on a final longitudinal distribution (\textless2\% mean difference between consecutive iterations). Figure~\ref{fig:iteration}(b) shows the different iterations in this process; the red line shows the longitudinal distribution produced from the initial model (Iteration 1), and the black line shows the longitudinal distribution used in the final version of the forward model (Iteration 6).

This new forward model is now used to retrieve the SO\textsubscript{2} column densities for all of the TEXES IRTF Io observations obtained since 2001. The full set of 191 retrieved SO\textsubscript{2} column densities is available in \cite{giles26_data}. We note that this new version of the forward model does still have limitations; it is based on data obtained at one specific season and when applying it to data obtained at other seasons, as in Section~\ref{sec:seasonal}, it implicitly assumes that the relative longitudinal distribution remains constant with season. We will show in Section~\ref{sec:results} that this is not true. Nevertheless, it captures the broad trends that are present at all seasons (higher density on the anti-Jovian hemisphere, lower density on the sub-Jovian hemisphere) and is a significant improvement over the flat distribution that was previously assumed in the forward model.

\subsection{Seasonal modeling}
\label{sec:seasonal}

\begin{figure*}
\centering
\includegraphics[width=13cm]{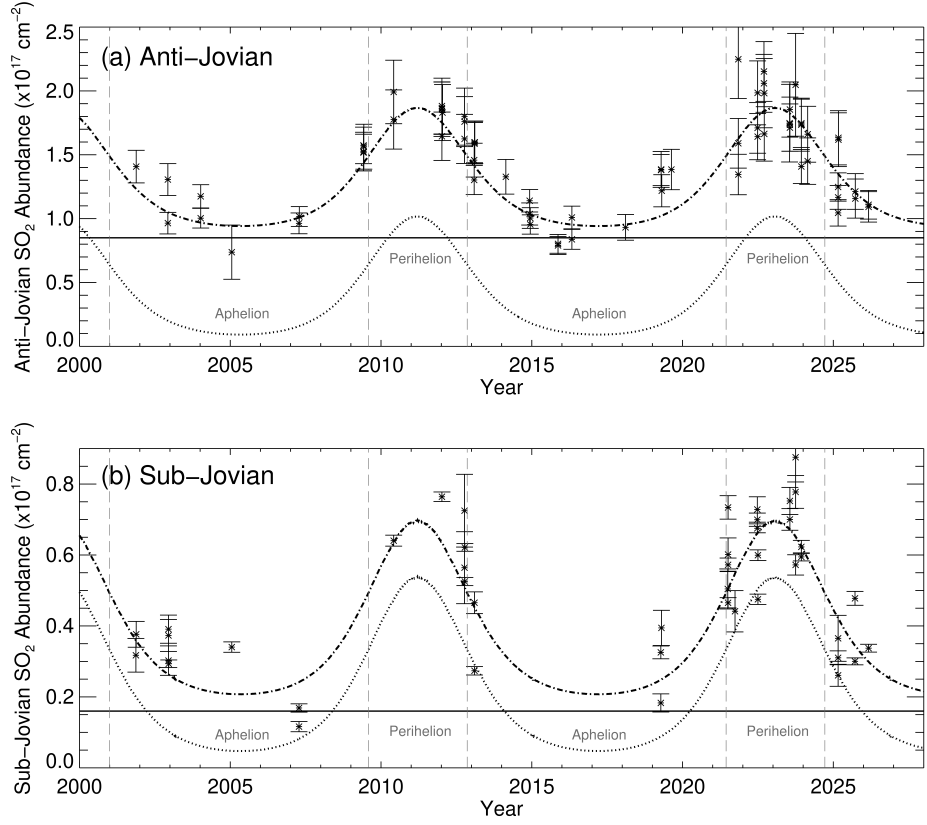}
\caption{The seasonal trends in the anti-Jovian and sub-Jovian equatorial SO\textsubscript{2} column abundances. The anti-Jovian data cover longitudes of 120--240$^{\circ}$W and are normalized to 180$^{\circ}$W. The sub-Jovian data cover longitudes of 300--60$^{\circ}$W and are normalized to 360$^{\circ}$W. For each hemisphere, the dash-dot curve shows the best-fit seasonal model, combining the SO\textsubscript{2} frost vapor-pressure equilibrium curve (dotted curve) and the constant, presumed volcanic, component (solid horizontal line).}
\label{fig:seasonal}
\end{figure*}

We now repeat the seasonal model analysis from~\cite{giles24}, using the updated SO\textsubscript{2} column densities and the additional data that have been obtained in the past two years. First, we sub-select data obtained from the anti-Jovian (120--240$^{\circ}$W) and sub-Jovian (300--60$^{\circ}$W) hemispheres and normalize each set of data to 180$^{\circ}$W and 360$^{\circ}$W respectively using the final iteration of the longitude distribution shown in Figure~\ref{fig:iteration}(b). These normalized data are shown in Figure~\ref{fig:seasonal}; while the precise values of the datapoints are different from~\cite{giles24} due to the changes made to the model, the qualitative temporal trend is very similar, with higher densities at perihelion and lower densities at aphelion. 

As in~\cite{giles24}, these data are used to constrain a seasonal model for each hemisphere. In order to find the best-fit seasonal model to the data, a one-dimensional numerical thermal model~\citep{spencer89,howett11} is used to calculate the frost surface temperature as a function of time for a range of Bond albedo and thermal inertia values; this model includes Jupiter eclipses whose effect varies with longitude and thus has different results for the anti-Jovian and sub-Jovian hemispheres. These surface temperatures are then converted into subsolar SO\textsubscript{2} column densities using the frost vapor-pressure sublimation equation described in~\cite{wagman79}. We then add a temporally constant, presumed volcanic, component, also covering a range of values. For each of the different combinations of the Bond albedo, thermal inertia and constant component, we calculate the reduced chi-square statistic, $\chi_{\nu}^2$; the best-fit solution is the one where $\chi_{\nu}^2$ is minimized. 

The two best-fit seasonal models for the anti-Jovian and the sub-Jovian hemispheres are shown by the dash-dot curves in Figure~\ref{fig:seasonal}. In each case, we also show the frost vapor-pressure equilibrium curve (dotted curve) and the constant component (solid horizontal line) that are added together to produce the best-fit model. Following the same error analysis process as~\cite{giles24}, we obtain the following best-fit parameters for the anti-Jovian/sub-Jovian hemispheres respectively: a bolometric Bond albedo of 0.66$^{+0.04}_{-0.11}$ / 0.63$^{+0.05}_{-0.12}$, a thermal inertia of 90$^{+190}_{-30}$ / 90$^{+190}_{-30}$ J m\textsuperscript{-2} s\textsuperscript{-1/2}K\textsuperscript{-1} (henceforth ``MKS'') and a constant component of 0.83$^{+0.04}_{-0.04}\times$10\textsuperscript{17} / 0.14$^{+0.04}_{-0.04}\times$10\textsuperscript{17} cm\textsuperscript{-2}. These values are very similar to the parameters fitted in~\cite{giles24}: a bolometric Bond albedo of 0.56$^{+0.04}_{-0.03}$ / 0.66$^{+0.03}_{-0.16}$, a thermal inertia of 250$^{+100}_{-90}$ / 80$^{+420}_{-20}$ MKS and a constant component of 0.74$^{+0.09}_{-0.11}\times$10\textsuperscript{17} / 0.06$^{+0.02}_{-0.06}\times$10\textsuperscript{17} cm\textsuperscript{-2}. As noted in~\cite{giles24}, they are also consistent with results from~\cite{walker12} who used HST/STIS and Galileo/PPR observations to infer a  Bond albedo of 0.55$\pm$0.02 and a thermal inertia of 200$\pm$50 MKS. Overall, we find that the changes made to the spectral modeling did not have a significant impact on the seasonal trends, and the seasonal trend has continued as expected in 2023--2026.

\subsection{Longitudinal distribution}
\label{sec:longitudinal}

In Section~\ref{sec:model_improvement}, where we first considered the longitudinal distribution of SO\textsubscript{2}, we restricted ourselves to data obtained at approximately the same heliocentric distance in order to avoid the additional seasonal variability. However, to obtain complete longitudinal curves with the finite number of observations we have available, we would ideally take advantage of the entire dataset we have, rather than just a small subset. 

In the previous section and in our previous paper~\citep{giles24}, we used an assumed longitudinal distribution to normalize data to a given longitude in order to study the seasonal trend. Now, we do the inverse; we use the seasonal curves shown in Figure~\ref{fig:seasonal} to normalize data to a given season in order to study the longitudinal distribution. In the previous analysis, we produced separate seasonal curves for the anti-Jovian and sub-Jovian hemispheres. Similarly, here we produce separate longitudinal distributions for the perihelion and aphelion time periods.

To achieve this, we first define a season range using the best-fit curves in Figure~\ref{fig:seasonal}; the season boundaries are shown by the vertical dashed lines. For perihelion, we include all data obtained on dates with a heliocentric distance of less than 5.04 AU and for aphelion, we include all data obtained on dates with a heliocentric distance of greater than 5.04 AU. 5.04 AU is closer to perihelion than aphelion, but was chosen as the dividing point because the rate of change of atmospheric density is greater near perihelion than near aphelion, as seen in Figure~\ref{fig:seasonal}. However, we note that the results are not sensitive to the precise boundary value used. We then use the two seasonal curves shown in Figure~\ref{fig:seasonal} to normalize the near-perihelion data to perihelion (the highest SO\textsubscript{2} density) and to normalize the near-aphelion data to aphelion (the lowest SO\textsubscript{2} density). The two seasonal curves from the two hemispheres each produce different normalization factors (there is larger relative variability on the sub-Jovian hemisphere), so for each data point, we use the central longitude to interpolate between the anti-Jovian curve and the sub-Jovian curve. The final results of this analysis are presented in Section~\ref{sec:results} and the effects of the normalization step are shown in Appendix~\ref{sec:appendix}.

We note that the seasonal trend and the longitudinal trend are not independent, as the seasonal model is constructed using a longitude normalization derived from the initial longitudinal analysis (Figure~\ref{fig:iteration}), while the final longitudinal distributions are subsequently obtained using the seasonal model. However, we find that repeating the steps described in Sections~\ref{sec:seasonal} and~\ref{sec:longitudinal} using instead the longitudinal distributions shown in Figure~\ref{fig:longitudinal_distribution} has a negligible impact on the final longitudinal distributions, demonstrating that we are not sensitive to the coupling between these two analyses.

We also note that, because our observed are not spatially resolved, we must assume a latitudinal distribution and we apply the same ``modified latitude'' relative distribution from~\cite{spencer05}, approximating the latitude distribution observed by~\cite{jessup04} at all longitudes. The SO\textsubscript{2} column densities that we present in Section~\ref{sec:results} are the equatorial column densities, assuming that relative latitudinal distribution. If there are, in reality, significant variations in the relative latitudinal distribution, then this would affect our inferred longitudinal distribution; for example, if the column density drops off more sharply with latitude at a given longitude, a higher equatorial column density would be required to reproduce the overall disk-averaged spectrum. However, we expect this effect to be small given that the flux from Io is dominated by the low-latitudes and we therefore expect our longitudinal SO\textsubscript{2} distributions to be representative of variations in the low-latitude atmosphere.

\section{Results and Discussion}
\label{sec:results}

\subsection{Perihelion and aphelion longitudinal distributions}

\begin{figure*}
\centering
\includegraphics[width=11cm]{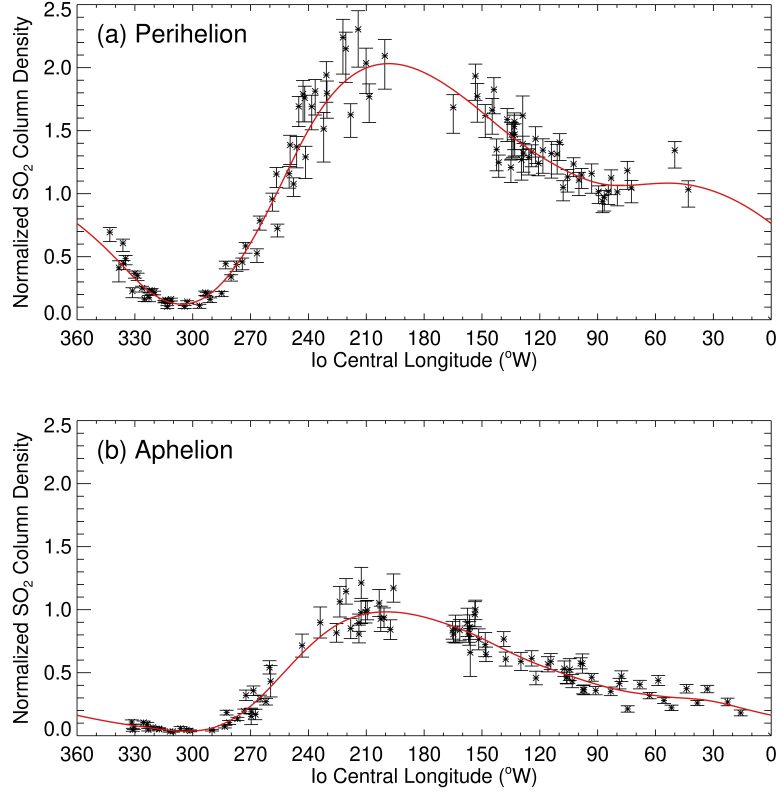}
\caption{The longitudinal distribution of the equatorial SO\textsubscript{2} column abundances. The perihelion data cover dates with a heliocentric distance of less than 5.04 AU and are normalized to perihelion. The aphelion data cover dates with a heliocentric distance of greater than 5.04 AU and are normalized to aphelion. In each case, the red line shows a cubic spline fit to the data points.}
\label{fig:longitudinal_distribution}
\end{figure*}

The longitudinal distribution of the equatorial SO\textsubscript{2} column abundances is shown in Figure~\ref{fig:longitudinal_distribution}. Figure~\ref{fig:longitudinal_distribution}(a) shows IRTF/TEXES data obtained near-perihelion and Figure~\ref{fig:longitudinal_distribution}(b) shows IRTF/TEXES data obtained near-aphelion. The data were normalized to perihelion and aphelion respectively, as described in Section~\ref{sec:longitudinal}; as seen in Figure~\ref{fig:seasonal}, the overall densities are lower at aphelion than at perihelion. The red lines show cubic spline fits to the data points.

In both cases, we see the broad longitudinal trend that has previously been reported: an SO\textsubscript{2} atmospheric density that peaks on the anti-Jovian hemisphere of Io, with column densities an order of magnitude higher than at sub-Jovian longitudes~\citep{jessup04,spencer05,feaga09,tsang13,giles24}. By combining data from many observing runs, considering the seasonal effects, and using our improved model of the longitudinal distribution, we are able to present a more complete picture of the longitudinal distribution than previous mid-infrared analyses.

Figure~\ref{fig:longitudinal_distribution} shows that the cubic spline fits to the perihelion and aphelion have similar shapes, with maxima near 200$^{\circ}$W and minima near 305$^{\circ}$W. Both longitudinal fitted curves also have a sharper dropoff on Io's trailing hemisphere (200--300$^{\circ}$W) than on the leading hemisphere (0--180$^{\circ}$W). The primary difference between the two fits is the magnitude of the contrast between the maximum and the minimum. At perihelion, the maximum SO\textsubscript{2} density is 16 times higher than the minimum density. At aphelion, the contrast is a factor of 27. However, we consider this inferred difference in contrast provisional, given the highly non-uniform sampling of longitude and season, and the normalizations required to generate the curves. We note that the hemispheric asymmetries we report here are much greater than the factor of 5 contrast reported by~\cite{feaga09}, but this is unsurprising given the findings by ~\cite{giono21} that Ly-\textalpha~imaging is not sensitive to the highest SO\textsubscript{2} abundances present at the equator.

In the following sections, we discuss the different mechanisms that could produce this longitudinal inhomogeneity and we compare our SO\textsubscript{2} density distributions to other published longitudinal results. In Figure~\ref{fig:longitudinal_comparisons}, our IRTF/TEXES perihelion and aphelion SO\textsubscript{2} distributions are shown in black (solid line and dotted line respectively) alongside longitudinal distributions of volcanic activity and SO\textsubscript{2} surface ice and the expected longitudinal distribution due to eclipses. In each case, the datasets have been normalized to a peak value of 1 to better visualize any similarities and differences in the shapes of the distributions.

\subsection{Comparison with volcanic hotspots}

\begin{figure*}
\centering
\includegraphics[width=16cm]{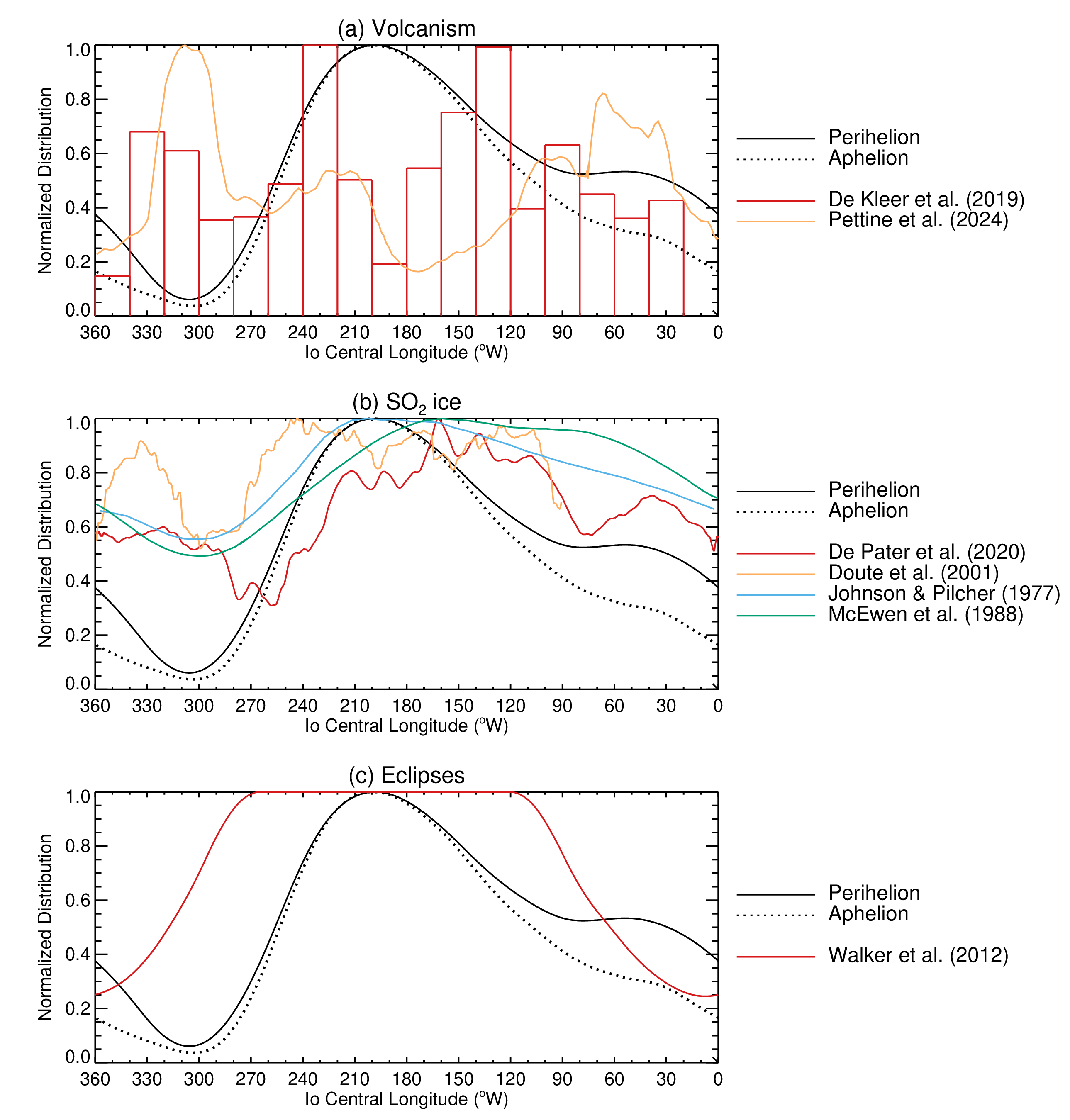}
\caption{A comparison of the longitudinal distribution of equatorial SO\textsubscript{2} column abundances with various other Io longitudinal datasets. All datasets have been normalized to a peak value of 1. (a) A comparison with two datasets showing the distribution of Io's volcanic hotspots. (b) A comparison with four datasets showing the distribution of SO\textsubscript{2} surface ice. (c) A comparison with a model that takes into account both eclipses and surface frost variability, with the eclipse effect dominating. See text for details.}
\label{fig:longitudinal_comparisons}
\end{figure*}

While there is increasing evidence that Io's atmosphere is dominantly supported by sublimation~\citep{jessup04,saur04,retherford07,tsang12,lellouch15,tsang16,giles24}, our model fits also require a seasonally-constant component which we have attributed to volcanism (Figure~\ref{fig:seasonal}).  If volcanic supply is important, then the spatial distribution of SO\textsubscript{2} gas in Io's atmosphere is likely to be affected by the geographical distribution of volcanoes across Io's surface. \cite{feaga09} suggested that this was the primary cause of the difference in the atmospheric density between the sub-Jovian and anti-Jovian hemispheres.

\cite{dekleer19} tracked Io's volcanic hotspots over a 5-year period using 2-5 \textmu m imaging with adaptive optics on the Keck and Gemini North telescopes. They used these data to measure the distribution of detected hotspots as a function of longitude and found that the number density of hotspots is highest on the leading and trailing hemispheres and that there is lower activity near the sub-Jovian and anti-Jovian longitudes. We show their data in red in Figure~\ref{fig:longitudinal_comparisons}(a), normalized to a peak value of 1 to allow easy comparison. Another longitudinal distribution was derived by~\cite{pettine24}, using infrared observations from the Juno spacecraft to map the volcanic flux on Io. Their (normalized) results are shown in orange in Figure~\ref{fig:longitudinal_comparisons}(a); while this volcanic flux distribution is not particularly similar to the hotspot number density distribution shown in red, they do both show an area of low volcanic activity around 180$^{\circ}$W. Neither of these volcanic distributions correlate well with our IRTF/TEXES SO\textsubscript{2} longitudinal distributions (black solid and dotted curves), which peak near 180$^{\circ}$W. This lack of correlation suggests that either the longitudinal asymmetry we observe in Io's atmospheric density is not primarily driven by volcanic sources, or that infrared volcanic thermal emission is not a good proxy for volcanic SO\textsubscript{2} gas emission.

\subsection{Comparison with frost distribution}

If Io's atmosphere is supported by sublimation, then a proposed driver of the longitudinal asymmetry in the atmosphere is spatial variability of the SO\textsubscript{2} surface ice~\citep{jessup04,spencer05,depater21}. In Figure~\ref{fig:longitudinal_comparisons}(b), we compare our SO\textsubscript{2} longitudinal distributions with the results of four different SO\textsubscript{2} surface-ice maps~\citep{depater20b,doute01,johnson77,mcewen88}.

\cite{depater20b} used ground-based observations from the Keck telescopes to develop an SO\textsubscript{2} ice map in the form of the equivalent width of the 2.12 \textmu m absorption band. This method of mapping the SO\textsubscript{2} ice is sensitive to only coarse-grained SO\textsubscript{2} frost. They find that this coarse frost is most abundant at equatorial latitudes, with highest equatorial abundances at 120-180$^{\circ}$W and lowest equatorial abundances at 240-290$^{\circ}$W. We show their longitudinal distribution (averaged between 200$^{\circ}$N and 20$^{\circ}$S and normalized) in red in Figure~\ref{fig:longitudinal_comparisons}(b).

\cite{doute01} used the Galileo NIMS instrument to map the SO\textsubscript{2} ice. They made use of the full NIMS spectrum, including the strong 4.07 \textmu m absorption band, which is sensitive to very small grains. They found that the frost coverage is highest in the mid-latitudes on the anti-Jovian hemisphere, with the minimum around 270-300$^{\circ}$W. Their longitudinal distribution (averaged between 200$^{\circ}$N and 20$^{\circ}$S and normalized) is shown in orange in Figure~\ref{fig:longitudinal_comparisons}(b).

Figure~\ref{fig:longitudinal_comparisons}(b) also shows two disk-averaged longitudinal profiles for near-UV (NUV) bright SO\textsubscript{2} frost, in blue and green. ~\cite{mcewen88} used Voyager multispectral mosaics to map the fractional coverage of NUV-bright SO\textsubscript{2} frost, and also calculated a disk-averaged version, which we normalize and show in green in Figure~\ref{fig:longitudinal_comparisons}(b). The blue curve, from~\cite{johnson77}, shows the (normalized) disk-integrated 350/550 nm brightness ratio, derived from ground-based observations. This ratio also acts as a proxy for the distribution of NUV-bright frost.

All four of these SO\textsubscript{2} frost distributions show similarities with the longitudinal distribution of atmospheric SO\textsubscript{2} obtained from our IRTF/TEXES data. All have the highest frost coverage on the anti-Jovian hemisphere, near to where the SO\textsubscript{2} column density also peaks at $\sim$200$^{\circ}$W, and the lowest on the sub-Jovian hemisphere, where the SO\textsubscript{2} column density has a minimum at $\sim$305$^{\circ}$W. They also all exhibit a sharper drop off in the frost coverage on the trailing hemisphere (270$^{\circ}$W) than on the leading hemisphere (90$^{\circ}$W), which also matches the SO\textsubscript{2} column density distribution. The disk-integrated frost distribution from~\cite{johnson77} is particularly similar to the SO\textsubscript{2} column density distribution, with the maxima and minima coinciding almost perfectly and an extremely similar overall shape. 

As noted in~\cite{giles24}, surface fractional frost coverage does not affect column density in simple vapor pressure equilibrium models.   In equilibrium, local upward sublimation flux from frost patches is balanced by the downward flux of molecules impacting and sticking to those patches from the atmosphere, which is a function of atmospheric density.  This is true regardless of the areal coverage of the frost patches. In these models, the only frost parameter that controls the column density is the frost temperature. However, these simple models are likely not an accurate representation of Io's complex atmosphere-surface interactions. \cite{walker10} developed a three-dimensional rarefied gas dynamics model of Io's sublimation-driven atmosphere and \cite{gratiy10} used this model to study how the frost distribution can drive the observed hemispheric asymmetry in SO\textsubscript{2} density. They found that the fractional frost coverage can significantly affect the SO\textsubscript{2} column density if the residence times for SO\textsubscript{2} molecules landing on frost-free surfaces are long (5000 seconds), which is what is expected for a highly porous surface. Using the fractional frost map of \cite{doute01}, \cite{gratiy10} showed that variations in the fractional frost coverage alone can explain a factor of two difference in SO\textsubscript{2} density between the anti-Jovian hemisphere and the sub-Jovian hemisphere. While this is smaller than the factor of 16.4 difference shown in Figure~\ref{fig:longitudinal_distribution}(a) and the the factor of 27.1 difference shown in Figure~\ref{fig:longitudinal_distribution}(b), the frost distribution is likely a significant contributing factor to the atmospheric variability. 

\subsection{Effect of eclipses}

\cite{tsang12} and \cite{walker12} proposed that the lower SO\textsubscript{2} column density at sub-Jovian latitudes could be due to the regular eclipses by Jupiter that are experienced by that hemisphere. The loss of sunlight for 2 hours per day acts to lower the diurnally-averaged surface temperature, and the small amount of additional thermal radiation that the sub-Jovian hemisphere receives from Jupiter itself is not enough to make up for this. The lower temperatures decrease the sublimation of the surface ice and lead to lower atmospheric densities.

\cite{walker12} carried out global simulations of Io's atmosphere using a direct simulation Monte Carlo (DSMC) method. Their model includes inhomogeneous frost distribution based on~\cite{doute01}. They analyzed in detail the effect that eclipses have on Io's atmospheric winds, temperatures and densities. They find that during eclipse, the dayside atmosphere near the sub-Jovian point becomes so thin that high energy ions from the Jovian plasma torus reach the surface, and this energy deposit alters the atmospheric thermal profile. The condensing atmosphere weakens the day-night pressure gradient, which changes the circumplanetary flow from supersonic to subsonic and disrupts the oblique shock near the dawn terminator. After eclipse, the atmosphere gradually returns to its former quasi-steady state. They find that the anti-Jovian point has a peak column density 4.2 times larger than the sub-Jovian point, which they attribute largely to the eclipse effect. A similar result is found by \cite{dott25}, who developed a time-dependent surface temperature model that assumed a constant distribution of frost across the surface, and explored what impact the eclipses alone can produce on the atmospheric densities. They find that the eclipses experienced by the sub-Jovian hemisphere can lead to maximum column densities that are 4 times lower than on the anti-Jovian hemisphere. Although a factor of 4 difference is not enough to explain by itself the longitudinal variability seen in our dataset, eclipses likely do play a major role.

As part of their analysis, \cite{walker12} produced an averaged dayside atmospheric density map for direct comparison with Ly-\textalpha~imaging maps. For this model, they simulated an Io orbit in the absence of collisions as a fully collisional model across all longitudes was too computationally expensive. The red line in Figure~\ref{fig:longitudinal_comparisons}(c) is adapted from the equatorial results of their model. This profile demonstrates the impact that eclipses have on the atmospheric density as a function of longitude. The predicted density is lowest near the sub-Jovian point and increases towards the anti-Jovian hemisphere, in agreement with the longitudinal profile obtained from the IRTF/TEXES data. However, the \cite{walker12} model then remains almost constant across the longitudes on the anti-Jovian hemisphere that do not directly experience the eclipse. Eclipses alone cannot account for the considerable longitudinal variability of atmospheric density on the anti-Jovian hemisphere, or the atmospheric asymmetry seen between the leading and trailing hemispheres. This is also noted by \cite{walker12}, who suggest that SO\textsubscript{2} ice inhomogeneity may play a larger role than their model currently assumes. 

\section{Conclusions}
\label{sec:conclusions}

In this paper, we use 25 years of IRTF/TEXES observations to study the longitudinal inhomogeneity of Io's  dayside atmosphere. The high spectral resolution observations are centered on mid-infrared SO\textsubscript{2} absorption lines and allow the SO\textsubscript{2} column density to be retrieved. By obtaining a total of 191 disk-averaged observations, each with a different sub-observer central longitude, we are able to study the longitudinal variability of Io's atmosphere in more depth than has previously been possible.

We use the complete long-term dataset to update the seasonal model previously fitted in~\cite{giles24}. We then use this seasonal model to apply a temporal normalization to the retrieved SO\textsubscript{2} densities, allowing us to directly compare observations obtained at slightly different seasons. Using this method, we produce two complete longitudinal distributions of Io's equatorial SO\textsubscript{2} column abundances, one representing perihelion and one representing aphelion. Both show the same hemispheric asymmetry that has been observed in previous studies~\citep{jessup04,spencer05,feaga09,tsang13,giles24}, with highest densities on the anti-Jovian hemisphere near 190$^{\circ}$W and lowest densities on the sub-Jovian hemisphere near 305$^{\circ}$W. While the overall densities are lower at aphelion than at perihelion (as discussed in~\cite{tsang12} and~\cite{giles24}), the contrast between the anti-Jovian and the sub-Jovian hemisphere appears to be greater at aphelion: a factor of $\sim$27 at aphelion and a factor of $\sim$16 at perihelion. We find that both longitudinal distributions have a sharper dropoff on Io's trailing hemisphere than on the leading hemisphere. 

We then compare these longitudinal distributions to the distributions expected if the spatial variability were primarily driven by volcanic activity, SO\textsubscript{2}-ice inhomogeneity or the effects of eclipses. Longitudinal distributions of volcanic activity from~\cite{dekleer19} and~\cite{pettine24} both show local minima in volcanic activity near 180$^{\circ}$W and do not correlate well with the atmospheric density distributions, suggesting that either volcanic activity does not drive Io's atmospheric inhomogeneity, or that infrared volcanic thermal emission is not a good proxy for volcanic SO\textsubscript{2} gas emission.

 In contrast, the SO\textsubscript{2}-ice distributions obtained from~\cite{depater20b},~\cite{doute01},~\cite{johnson77} and~\cite{mcewen88} all show similarities with the atmospheric density distributions, with maxima on the anti-Jovian hemisphere, minima on the sub-Jovian hemisphere and sharper dropoffs on the trailing hemisphere than the leading hemisphere. The effect of eclipses also produces significant hemispheric asymmetries~\citep{walker12,dott25} but is insufficient to explain the observed longitudinal variability within the non-eclipsed anti-Jovian hemisphere. We therefore conclude that the longitudinal distribution of Io's dayside atmospheric density is primarily driven by a combination of the eclipse effect and the inhomogenity of the SO\textsubscript{2} surface ice, with volcanic activity playing a more minor role.

\section*{Data Availability}

The reduced IRTF/TEXES data are available at the NASA/IPAC Infrared Science Archive. The full set of retrieved SO\textsubscript{2} column densities and the data to generate each figure in this paper are available in~\cite{giles26_data}. 

\begin{acknowledgments}

The authors were Visiting Astronomers at the Infrared Telescope Facility, which is operated by the University of Hawaii under contract 80HQTR19D0030 with the National Aeronautics and Space Administration. This work was funded by NASA Solar System Observations Grant 80NSSC21K1138. 

\end{acknowledgments}

\begin{contribution}

RSG performed the telescope observations, reduced the data, refined the spectral model and was responsible for writing and submitting the manuscript.
JRS came up with the initial research concept, developed the initial spectral and seasonal models, and edited the manuscript.
TKG performed the telescope observations. 
CCCT developed the initial spectral and seasonal models.
EL developed the initial spectral model.
MALV developed the initial spectral model.

\end{contribution}

\facilities{IRTF(TEXES)}

\newpage
\appendix

\section{Effect of new model and seasonal normalization}
\label{sec:appendix}

\begin{figure*}[h]
\centering
\includegraphics[width=15cm]{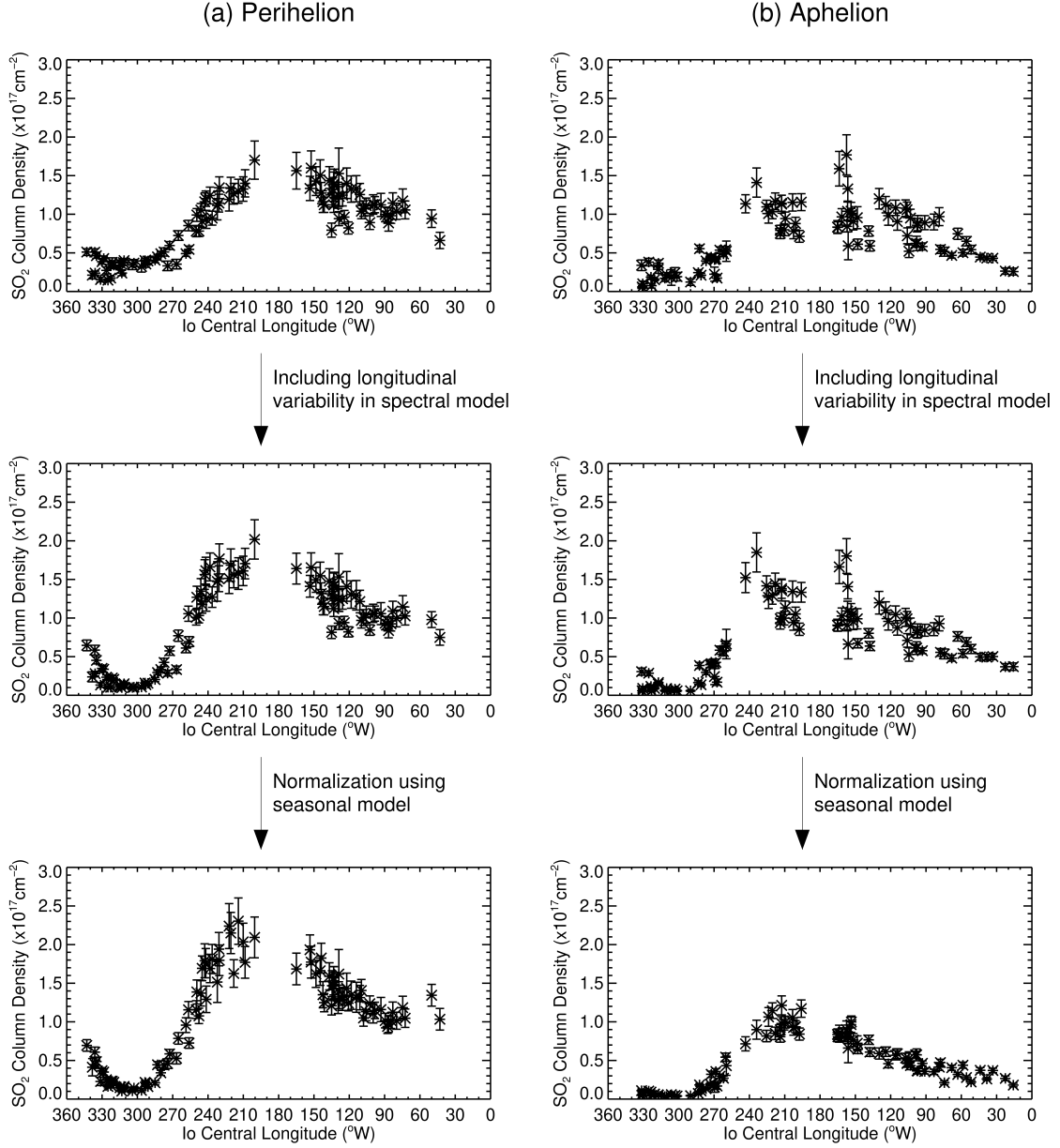}
\caption{Steps taken in the production of longitudinal distributions of Io's atmospheric density around (a) perihelion and (b) aphelion. The top panels show the perihelion and aphelion distributions using the initial spectral model, which assumed a constant equatorial density across Io's disk for a given sub-observer longitude. The middle panels show the distributions that are obtained using the updated spectral model, which incorporated longitudinal variability. The bottom panels how the distributions that are obtained once the data are temporally normalized to perihelion and aphelion respectively. As expected, this normalization step decreases the scatter in the data points and leads to a smoother curve.}
\label{fig:normalization_effect}
\end{figure*}

\newpage
\bibliographystyle{aasjournalv7}

\end{document}